\documentclass[epsf]{elsart}
\usepackage{amssymb}
\usepackage{epsf,amsfonts,amssymb}
\usepackage{amsmath}
\usepackage{url,epsfig}

\begin{document}

\begin{frontmatter}
\title{Nonlinear oscillations and stability domains in fractional reaction-diffusion systems}
\author[ima]{V.Gafiychuk},
\author[imath]{B.Datsko},
\author[imath]{V.Meleshko}

\address[ima]{Institute of Computer Modeling, Krakow University of Technology, 24
Warszawska Street,  Krakow, Poland, 31155}
\address[imath]{Institute of Applied Problem of Mechanics and Mathematics, National
Academy of Sciences of Ukraine, Naukova Street 3 B,  Lviv,
Ukraine, 79053}

\begin{abstract}
We study a fractional reaction-diffusion system with two types of
variables: activator and inhibitor. The interactions between
components are modeled by cubical nonlinearity. Linearization of
the system around the homogeneous state provides information about
the stability of the solutions which is quite different from
linear stability analysis of the regular system with integer
derivatives. It is shown that by combining the fractional
derivatives index with the ratio of characteristic times, it is
possible to find the marginal value of the index where the
oscillatory instability arises. The increase of the value of
fractional derivative index leads to the time periodic solutions.
The domains of existing of periodic solutions for different
parameters of the problem are obtained. A computer simulation of
the corresponding nonlinear fractional ordinary differential
equations is presented. It is established by computer simulation
that there exists a set of stable spatio-temporal structures of
the one-dimensional system under the Neumann and periodic boundary
condition. The features of these solutions consist in the
transformation of the steady state dissipative structures to
homogeneous oscillations or space temporary structures at a
certain value of fractional index and the ratio of characteristic
times of system.
\end{abstract}

\end{frontmatter}

\section{Introduction}

Reaction--diffusion systems (RDS) have been extensively used in
the study of self-organization phenomena in physics, biology,
chemistry, ecology etc. (See, for example
\cite{pr,ch,m90,KO,dk89,dk03,lg,gl,sbg,mo02}). The main result
obtained from these systems is that nonlinear phenomena include
diversity of stationary and spatio-temporary dissipative patterns,
oscillations, different types of waves, excitability, bistability
etc. The mechanism of the formation of such type of nonlinear
phenomena and the conditions of their emergence have been
extensively studied during the last couple decades.

In the recent years, there has been a great deal of interest in fractional
reaction-diffusion (FRD) systems \cite{hw,hw1,he02,add1,gd,GDI05,ar1,zw,vb}
which from one side exhibit self-organization phenomena and from the other
side introduce a new parameter to these systems, which is a fractional
derivative index, and it gives a great degree of freedom for diversity of
self-organization phenomena.

At the same time, analyzing structures in FRD systems evolves,
both from the point of view of the qualitative analysis and from
the computer simulation. Namely these two problems are the goals
of our investigation here. Our particular interest is the analysis
of the specific non-linear system of FRD equations. We consider a
very well-known example of the RDS with cubical nonlinearity
\cite{m90,KO,lg} which probably is the simplest one used in RD
systems modeling
\begin{equation}
\tau _{1}\frac{\partial ^{\alpha }n_{1}(x,t)}{\partial t^{\alpha }}%
=l^{2}\nabla ^{2}n_{1}(x,t)+n_{1}-n_{1}^{3}/3-n_{2},  \label{1}
\end{equation}%
\begin{equation}
\tau _{2}\frac{\partial ^{\alpha }n_{2}(x,t)}{\partial t^{\alpha }}%
=L^{2}\nabla ^{2}n_{2}(x,t)-n_{2}+\beta n_{1}+\mathcal{A},
\label{2}
\end{equation}%
subject to

(i) Neumann: \
\begin{equation}
dn_{i}/dx|_{x=0}=dn_{i}/dx|_{x=l_{x}}=0,i=1,2.  \label{bc1}
\end{equation}%
or

(ii) Periodic:
\begin{equation}
n_{i}(t,0)=n_{i}(t,l_{x}),\;dn_{i}/dx|_{x=0}=dn_{i}/dx|_{x=l_{x}},i=1,2.
\label{bc2}
\end{equation}%
boundary conditions and with the certain initial condition
$n_{i}|_{t=0}=n_{i}^{0}(x)$. Here $x:0\leq x\leq l_{x};(x,t)\in $ $\mathbb{%
R\times R}_{+}$; $\nabla ^{2}=\frac{\partial ^{2}}{\partial x^{2}}%
;n_{1}(x,t) $, $n_{2}(x,t)\in $ $\mathbb{R}$ -- activator and
inhibitor variables correspondingly, $\tau _{1},\tau _{2},l,L,\in $ $\mathbb{R}$ --
characteristic times and lengths of the system, $\mathcal{A}\in $
$\mathbb{R}$ -- is an external parameter.

Fractional derivatives $\frac{\partial ^{\alpha }n(x,t)}{\partial t^{\alpha }%
}$ on the left hand side of equations (\ref{1}),(\ref{2}) instead of
standard time derivatives are the Caputo fractional derivatives in time of
the order $0<\alpha <2$ and are represented as \cite{skm,po}

\begin{equation*}
\frac{\partial ^{\alpha }}{\partial t^{\alpha }}n(t):=\frac{1}{\Gamma
(m-\alpha )}\int\limits_{0}^{t}\frac{n^{(m)}(\tau )}{(t-\tau )^{\alpha +1-m}}%
d\tau ,\mbox{}\;m-1<\alpha <m,m\in N.
\end{equation*}

It should be noted that equations (\ref{1}),(\ref{2}) at $\alpha
=1$ correspond to standard RDS. At $\alpha <1,$ they describe
anomalous sub-diffusion and at $\alpha >1$~ -- anomalous
super-diffusion.

\section{Linear stability analysis}

Stability of the steady-state constant solutions of the system (\ref{1}),(%
\ref{2}) correspond to homogeneous equilibrium state
\begin{equation}
W=n_{1}-n_{1}^{3}/3-n_{2}=0, \quad Q=-n_{2}+\beta
n_{1}+\mathcal{A}=0 \label{e}
\end{equation}%
can be analyzed by linearization of the system nearby this
solution. In this case the system (\ref{1}),(\ref{2}) can be
transformed to linear system at equilibrium point (\ref{e}) and
linearized around this equilibrium state. As a result we have:
\begin{equation}
\frac{\partial ^{\alpha }\mathbf{u}(x,t)}{\partial t^{\alpha }}=\widehat{F}%
(u)\mathbf{u}(x,t),  \label{lin}
\end{equation}%
where $\mathbf{u}(x,t)=\left(
\begin{array}{c}
\triangle n_{1}(x,t) \\
\triangle n_{2}(x,t)%
\end{array}%
\right) ,$ $\widehat{F}(u)=\left(
\begin{array}{cc}
(l^{2}\nabla ^{2}+a_{11})/\tau _{1} & a_{12}/\tau _{1} \\
a_{21}/\tau _{2} & (L^{2}\nabla ^{2}+a_{22})/\tau _{2}%
\end{array}%
\right) $ -- is a Frechet derivative with respect to $\mathbf{u}(x,t),$ $%
a_{11}=W_{n_{1}}^{\prime },$ $a_{12}=W_{n_{2}}^{\prime },$ $%
a_{21}=Q_{n_{1}}^{\prime },$ $a_{22}=Q_{n_{2}}^{\prime }$ (all
derivatives are taken at homogeneous equilibrium states
(\ref{e})). By substituting the solution in the form
$\mathbf{u}(x,t)=\left(
\begin{array}{c}
\triangle n_{1}(t) \\
\triangle n_{2}(t)%
\end{array}%
\right) \cos kx,k=\frac{\pi }{l_{x}}j,j=1,2,...$ into FRD system (\ref{lin})
we can get the system of linear ordinary differential equations (\ref{lin})
with the matrix $F$ determined by the operator $\widehat{F}$.

For analyzing stability conditions of the equations (\ref{1}),(\ref{2}) let
us use simple linear transformation which can convert this linear system (%
\ref{lin}) to a diagonal form

\begin{equation}
\frac{d^{\alpha }\mathbf{\eta (}t\mathbf{)}}{dt^{\alpha }}=C\mathbf{\eta }%
(t),  \label{f}
\end{equation}%
where $C$ is a diagonal matrix for $F$: $\ C=P^{-1}FP=\left(
\begin{array}{cc}
\lambda _{1} & 0 \\
0 & \lambda _{2}%
\end{array}%
\right) ,$ eigenvalues $\lambda _{1,2}$ are determined by the characteristic
equation of the matrix $F$, $\lambda _{1,2}=\frac{1}{2}(trF\pm \sqrt{%
tr^{2}F-4\det F}),\ \mathbf{\eta }(t)=P^{-1}\left(
\begin{array}{c}
\triangle n_{1}(t) \\
\triangle n_{2}(t)%
\end{array}%
\right) ,$ $P$ \ is the matrix of eigenvectors of matrix $F$.

\begin{figure}[tbp]
\begin{center}
\begin{tabular}{cc}
\includegraphics[width=0.5\textwidth]{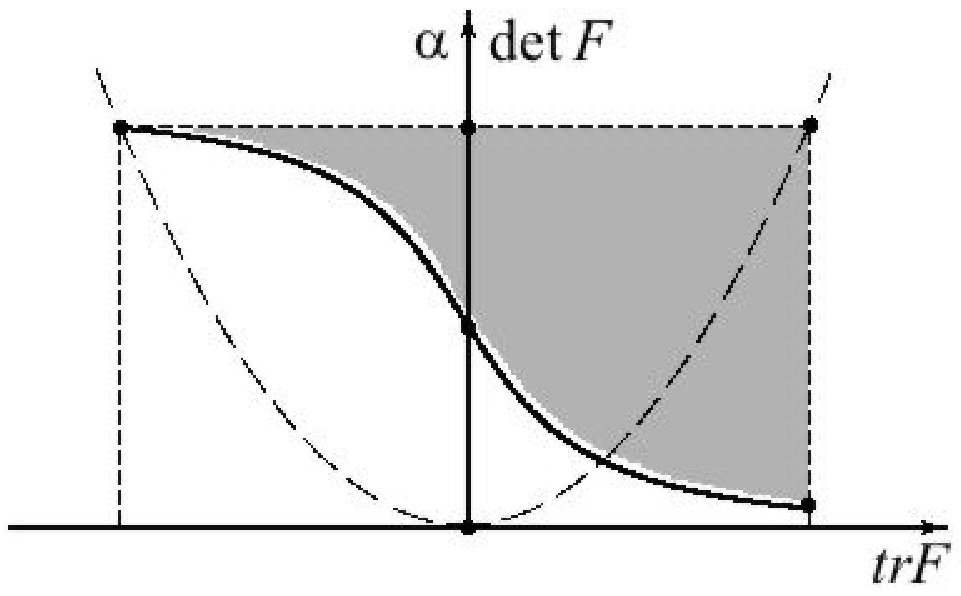} & %
\includegraphics[width=0.5\textwidth]{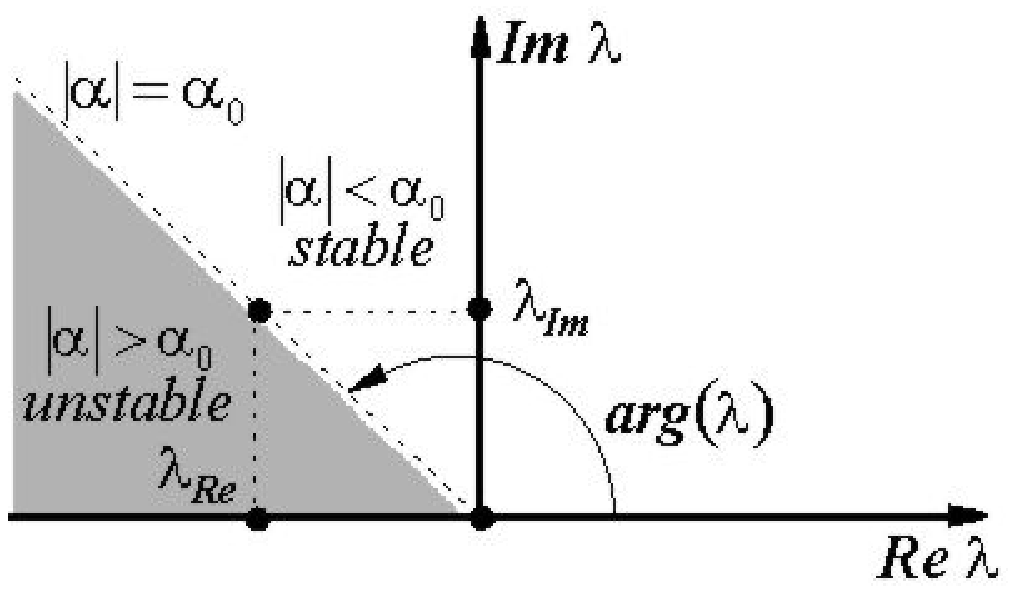} \\
(a) & (b)%
\end{tabular}%
\end{center}
\caption{Schematic view of the marginal curve of $\alpha$ (solid
line), describing fixed points for two-dimensional vector field --
(a), the position of eigenvalue $\lambda$ corresponding to
marginal value of $\protect\alpha$ in the coordinate system
($Re\lambda, Im\lambda$) -- (b). Shaded domains correspond to
instability region} \label{rys1}
\end{figure}

In this case, the solution of the vector equation (\ref{f}) is
given by Mittag-Leffler functions\bigskip\ \cite{tz,skm,po,os06,a}
\begin{equation}
\triangle n_{i}(t)=\sum\limits_{k=0}^{\infty }\frac{(\lambda _{i}t^{\alpha
})^{k}}{\Gamma (k\alpha +1)}\triangle n_{i}(0)=E_{\alpha }(\lambda
_{i}t^{\alpha })\triangle n_{i}(0),i=1,2.  \label{sol}
\end{equation}%
Using the result obtained in the papers \cite{GDI05,mi}, we can conclude
that if for any of the roots
\begin{equation}
|Arg(\lambda _{i})|<\alpha \pi /2  \label{100}
\end{equation}%
the solution has an increasing function component then the system is
asymptotically unstable.

Analyzing the roots of the characteristic equations, we can see that at $%
4\det F-tr^{2}F>0$ eigenvalues are complex and can be represented as
\begin{equation*}
\lambda _{1,2}=\frac{1}{2}(trF\pm i\sqrt{4\det F-tr^{2}F})\equiv \lambda
_{Re}\pm i\lambda _{Im}.
\end{equation*}%
The roots $\lambda _{1,2}$ are complex inside the parabola (Fig. \ref{rys1}%
(a)) and the fixed points are the spiral source ($trF>0$)\ \ or spiral sinks
($trF<0$). The plot of the marginal value $\alpha :\alpha =\alpha _{0}=$ $%
\frac{2}{\pi }|Arg(\lambda _{i})|$ which follows from the conditions (\ref%
{100}) is given by\ the formula
\begin{equation}
\alpha _{0}=\left\{
\begin{array}{cc}
\frac{2}{\pi }\arctan \sqrt{4\det F/tr^{2}F-1}, & trF\geq 0, \\
2-\frac{2}{\pi }\arctan \sqrt{4\det F/tr^{2}F-1}, & trF\leq 0%
\end{array}%
\right.  \label{al}
\end{equation}%
and is presented in the Fig. \ref{rys1}.

Let us analyze the system solution with the help of the Fig. \ref{rys1}%
(a). Consider the parameters which keep the system inside the parabola. It
is a well-known fact, that at $\alpha =1$\ the domain on the righthand
side of the parabola ($trF>0$) is unstable with the existing limit circle,
while the domain on the left hand side ($trF<0$) is stable. By crossing the
axis $trF=0$ the Hopf bifurcation conditions become true.

In the general case of \ $\alpha :0<\alpha <2$\ \thinspace\ for every point
inside the parabola there exists a marginal value of $\alpha _{0}$ where the
system changes its stability. The value of $\alpha $ is a certain
bifurcation parameter which switches the stable and unstable state of the
system. At lower $\alpha :$ $\alpha <\alpha _{0}=$ $\frac{2}{\pi }%
|Arg(\lambda _{i})|$ the system has oscillatory modes but they are stable.
Increasing the value of $\alpha >\alpha _{0}=$ $\frac{2}{\pi }|Arg(\lambda
_{i})|$ leads to instability. As a result, the domain below the curve\ $%
\alpha _{0}$, as a function of $trF$\thinspace\ is stable and the domain
above the curve is unstable.

The plot of the roots, describing the mechanism of the system instability,
can be understood from the Fig. \ref{rys1}(b) where the case $\alpha
_{0}>1 $ is described. In fact, having complex number $\lambda _{i}$ with $%
\mathit{Re}\lambda _{i}<0$\ at $\alpha \rightarrow 2$ it is always possible
to satisfy the condition $|Arg(\lambda _{i})|<\alpha \pi /2$, and the system
becomes unstable according to homogeneous oscillations (Fig \ref{rys1}%
(b)). The smaller is the value of $trF$, the easier it is to fulfill the
instability conditions.

In contrast to this case, a complex values of $\lambda _{i},$ with$\ \mathit{%
Re}\lambda _{i}>0$ lead to the system instability for regular system with $%
\alpha =1.$ However fractional derivatives with $\alpha <1$ can stabilize
the system if $\alpha <\alpha _{0}=$ $\frac{2}{\pi }|Arg(\lambda _{i})|.$
This makes it possible to conclude that fractional derivative equations with
$\alpha <1$\ are more stable that their integer twinges.

\section{Solutions of the coupled fractional ordinary differential equations
(FODEs)}

\begin{figure}[tbp]
\begin{center}
\begin{tabular}{cc}
\includegraphics[width=0.9\textwidth]{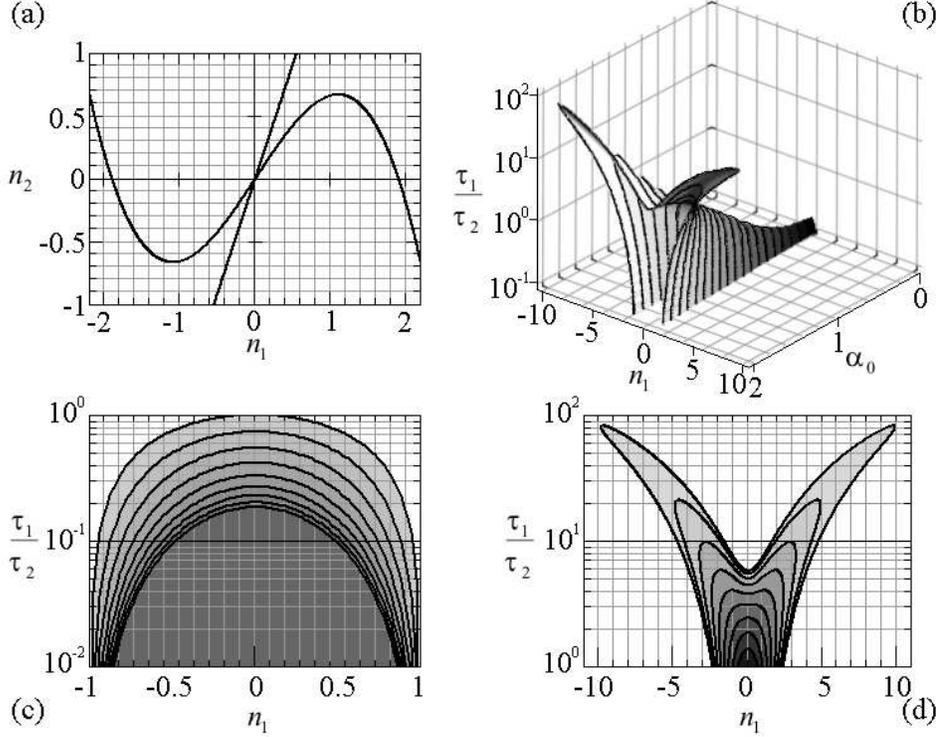} &
\end{tabular}%
\end{center}
\caption{Null isoclines -- (a), 3-d instability domains  in
coordinates $(\alpha_0,\overline{n}_{1},\protect\tau
_{1}/\protect\tau _{2})$ -- (b), dependence of $\protect\tau
_{1}/\protect\tau _{2}$ on $\overline{n}_{1}$ at $\alpha_0$
changing from 0.1 (bottom curve) to 1.0 (upper curve) with step
0.1 -- (c), dependence of $\protect\tau _{1}/\protect\tau _{2}$ on
$\overline{n}_{1}$ at $\alpha_0$ changing from 1.1 (bottom curve) to
1.9 (upper curve) with step 0.1 -- (d). The shaded domains
correspond to those one obtained by slicing 3-d surface
represented on figure (b).} \label{rys2}
\end{figure}

Let us first consider the coupled fractional ordinary differential
equations (FODEs) which can be obtained from (\ref{1}),(\ref{2})
at $l=L=0$ and analyze the stability conditions for such systems.
The plot of isoclines for this model is represented on Fig.
\ref{rys2}(a). In this case homogeneous solution can be determined
from the system of equations $W=Q=0$ and is given by the solution of
cubic algebraic equation
\begin{equation}
(\beta -1)\overline{n}_{1}+\overline{n}_{1}^{3}/3+\mathcal{A}=0.
\label{n1}
\end{equation}%
Simple calculation makes it possible to write useful expressions required
for our analysis

\begin{equation*}
F=-\left(
\begin{array}{cc}
(-1+\overline{n}_{1}^{2})/\tau _{1} & \quad 1/\tau _{1} \\
-\beta /\tau _{2} & \quad 1/\tau _{2}%
\end{array}%
\right) ,trF=\frac{(1-\overline{n}_{1}^{2})}{\tau _{1}}-\frac{1}{\tau _{2}}%
,\det F=\frac{(\beta -1)+\overline{n}_{1}^{2}}{\tau _{1}\tau _{2}}.
\end{equation*}

It is easy to see that if the value of $\tau _{1}/\tau _{2}$, in
certain cases, is smaller than $1$, the instability conditions
($trF>0$)\ lead to Hopf bifurcation for regular system ($\alpha
=1$) \cite{pr,ch,m90,KO,dk89}. In this case, the plot of the
domain, where instability exists, is shown on the Fig.
\ref{rys2}(c) (unstable domain is below the upper curve). The
linear analysis of the system for $\alpha =1$ shows that, if $\tau
_{1}/\tau _{2}>1$, the solution corresponding to the intersections
of two isoclines is stable. The smaller is the ratio of $\tau
_{1}/\tau _{2}$, the wider is the instability region. Formally, at
$\tau _{1}/\tau _{2}\rightarrow 0$, the
instability region in $\overline{n}_{1}$\ coincides with the interval ($-1,1$%
) where the null isocline $W(n_{1},n_{2})=0$ \ has its increasing part.
These results are very widely known in the theory of nonlinear dynamical
systems \cite{pr,ch,m90,KO,dk89}.

In the FODEs the conditions of the instability change (\ref{100}), and we
have to analyze the real and the imaginary part of the existing complex
eigenvalues, especially the equation:
\begin{equation}
4\det F-tr^{2}F=4((\beta -1)+\overline{n}_{1}^{2})/\tau _{1}\tau _{2}-\left(
(1-\overline{n}_{1}^{2})/\tau _{1}-1/\tau _{2}\right) ^{2}>0.  \label{ri}
\end{equation}%
In fact, with the complex eigenvalues, it is possible to find out the
corresponding value of $\alpha $ where the condition (\ref{100}) is true. We
will show that this interval is not correlated with the increasing part of
the null isocline of the system. Indeed, omitting simple calculation, we can
write an equation for marginal values of $\overline{n}_{1}$
\begin{equation}
\overline{n}_{1}^{4}-2(1+\frac{\tau _{1}}{\tau _{2}})\overline{n}_{1}^{2}+%
\frac{\tau _{1}^{2}}{\tau _{2}^{2}}-2\frac{\tau _{1}}{\tau _{2}}(2\beta
-1)+1=0,  \label{ri1}
\end{equation}%
and solution of this quadratic equation gives the domain where the
oscillatory instability arises:
\begin{equation}
\overline{n}_{1}^{2}=1+\frac{\tau _{1}}{\tau _{2}}\pm 2\sqrt{\beta \frac{%
\tau _{1}}{\tau _{2}}}  \label{t}
\end{equation}%
This expression estimates the maximum and minimum values of $\overline{n}%
_{1} $ where the system can be unstable at certain value of $\alpha =\alpha
_{0}$ as a function of $\tau _{1}/\tau _{2}$. On the Fig. 2 the dependance
$\tau _{1}/\tau _{2}$ is given as a function of $\overline{n}_{1}$ for
different values of $\alpha $ changing with the step $0.1$.

In fact, examine the domain of the FODEs where the eigenvalues are complex.
The condition (\ref{100}) determine dependence among values $\overline{n}%
_{1},\tau _{1}/\tau _{2}$ and $\alpha _{0}$ (this dependance is
represented on Fig. 2(b). On the Fig. 2(c) and 2(d)
cross sections of this figure is presented for fixed value of
$\alpha _{0}.$ Inside the curve system of FODEs is unstable and
outside it is stable. We can see that at $\alpha <\alpha _{0}$\
instability domain in coordinates \ ( $\overline{n}_{1},\tau
_{1}/\tau _{2}$) is smaller then for the case $\alpha _{0}=1$
(Fig. 2(c)). In the same time for the case $\alpha>\alpha _{0}$
instability domain increases and the greater is the value of $\tau
_{1}/\tau _{2}$ then the wider is the region in $\overline{n}_{1}$
where the instability holds true (Fig. 2(d)). In this case, from
(\ref{t})\ we immediately obtain approximate dependance
($\overline{n}_{1}\simeq \pm \sqrt{\tau _{1}/\tau _{2}}$). Of
course, at $\tau _{1}/\tau _{2},\overline{n}_{1}\gg 1$ the interval in $%
\overline{n}_{1}$ where the instability arises increases but the
domain decreases and gets the form of a narrow concave stripe
following parabola \ $\tau _{1}/\tau _{2}\simeq
\overline{n}_{1}^{2}$\ for $2-\alpha _{0}\ll 1.$

Finally, we can conclude that for $\tau _{1}/\tau _{2}>1$ in
contrast to the regular system with integer index $\alpha $\
fractional differential equations can be both stable and unstable.
The greater is the value of $\tau _{1}/\tau _{2}$ then the wider
is interval in $\overline{n}_{1}$ where\ instability conditions
are true. Opposite situation is for $\alpha <1$\ while regular
system with integer index is always unstable fractional dynamical
system can be stable. It is a statement, that FODEs are at least
as stable as their integer order counterparts \cite{po}. It is
obvious that this statement is true for $\alpha <1$ only.

It should be noted that, even if the eigenvalues are not complex ($\lambda
_{Im}=0$), the systems with fractional derivatives can poses oscillatory
damping oscillations. Such situation takes place when $4\det F-tr^{2}F<0,$ $%
trF<0,$ $\det F>0$ and two eigenvalues are real and less than zero. In this
case, at $1<\alpha <2$\ steady state solutions of the system are stable and
any perturbations are damping. Such system was considered, for example, in
the article \cite{zse05},\ where an analytical solution for fractional
oscillator is obtained.

Our task here is to confirm our linear analysis by finding out not
only the conditions of the bifurcation but also the real time
dynamics of FODEs by corresponding combination of the parameters $\tau _{1}/\tau _{2},\overline{n}%
_{1}$ for different values of $\alpha :$ $0<\alpha <2.$ We have
established here that the dynamics of the FODEs can be much more
complicated than that one of the equations with integer order
\cite{mi,El96,wz}.

\begin{figure}[tbp]
\begin{center}
\includegraphics[width=1.0\textwidth]{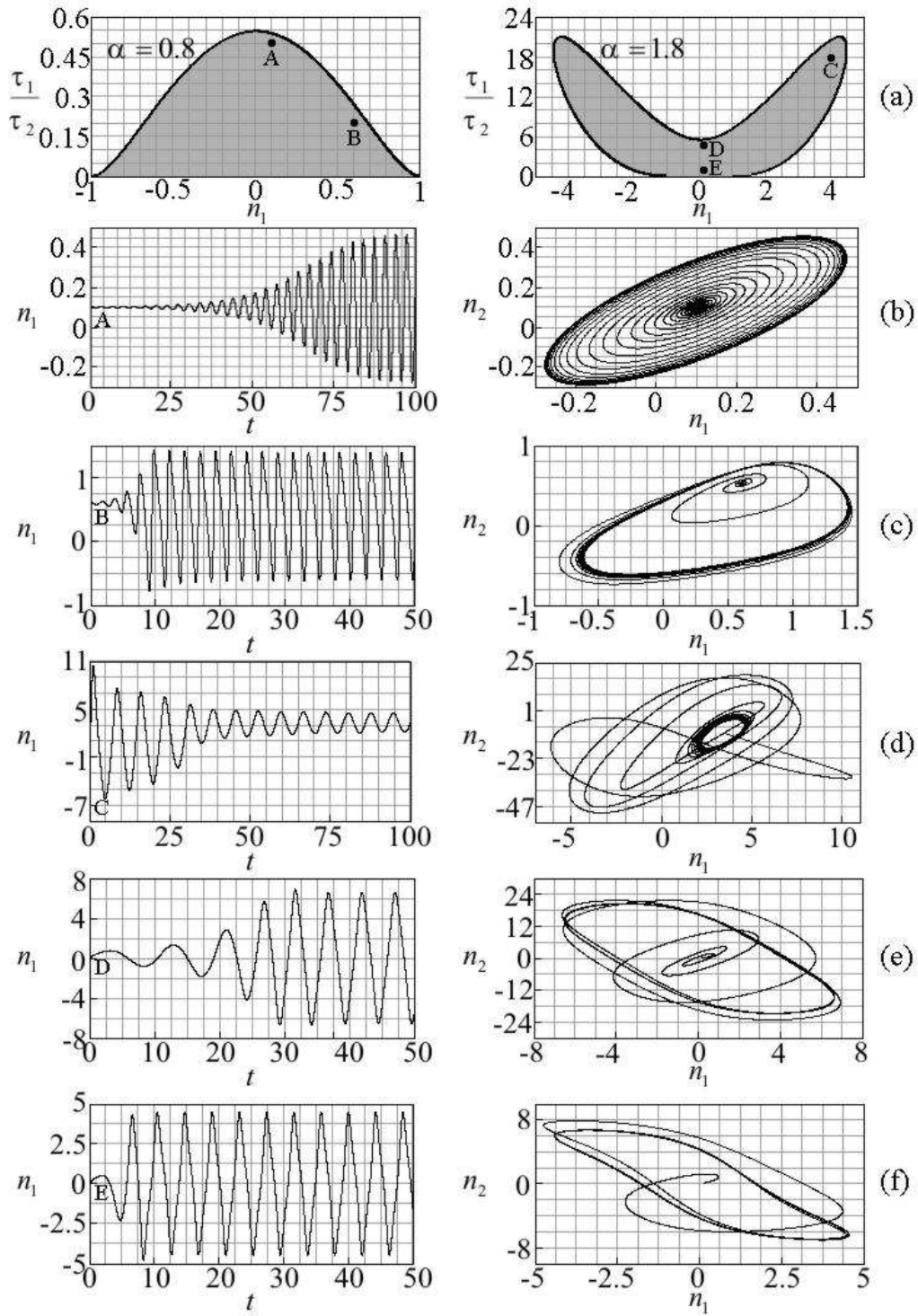}
\end{center}
\caption{Instability domains for $\alpha=0.8$ and $\alpha=1.8$ --
(a). The time domain oscillations (left) and corresponding two
dimensional phase portrait (right) for point A -- (b), B -- (c), C
-- (d), D -- (e) E -- (f) ($\tau_2=1,\protect\beta =2,
l=L=0$). For each point A, B, C, D and E parameter $\mathcal{A}$ corresponds
to steady state solution taken in these points.} \label{rys4}
\end{figure}

Fig. 3 gives the results of computer simulation of fractional
ODEs for different cases considered above. We single out two
different cases $\alpha <1$ and $\alpha >1$. The characteristic
instability domain (shaded region) for these two cases located on
the first row in two columns. Outside the
shaded region system is stable and solution is given by the equations (\ref%
{e}). For any point inside shaded region system is unstable and we
investigate nonlinear dynamics by computer simulation of the
FODEs. We present nonlinear dynamics for several points denoted by
capital letters A,B,C,D,E corresponding to homogeneous
distributions in these points (null isoclines intersect in these
points). For the points A and B taken from the certain domain
obtained at $\alpha=0.8 $ and for points C,D,E taken from  the
instability domain obtained at $\alpha $=1.8 nonlinear dynamics is
presented on Fig. 3(b)-(f).

From Fig. 3, we can tell that at the point A
increment of the oscillations is small enough and formation of the limit cycle is during long period ($%
t\simeq100$). At the point B oscillations develop more rapidly and
have a bigger amplitude. The result of computer simulation of the
system with $\tau _{1}/\tau _{2}>1$ is represented in points C, D,
E. In this case in point C oscillations have a small amplitude and
sufficiently large transient time. In the points D, E oscillations
get fast enough and have a great amplitude.

We presented simulation of the evolutionary dynamics for only two values of
indexes $\alpha$. Nevertheless, similar dynamics is inherent to any value of
$\alpha$ (we investigated the alpha from $\alpha$=0.1 till $\alpha$=1.9 with
step 0.1).

\section{Computer simulation of pattern formation}

This section contains a discussion of the results of the numerical
study of the initial value problem of the system
(\ref{1}),(\ref{2}). The system with corresponding initial and
boundary conditions was integrated numerically using the explicit
and implicit schemes with respect to time and centered difference
approximation for spatial derivatives. The fractional derivatives
were approximated using the scheme on the basis of
Grunwald-Letnikov definition for $0 < \alpha < 2$  \cite{po}. In
other words, for the system of $n$ fractional RD equations
\begin{equation}
\tau_j \frac{^C\partial^{\alpha_j} u_j(x,t)}{\partial t^{\alpha_j}}=d_j\frac{%
\partial^2 u_j(x,t)}{\partial x^2}+f_j(u_1,...,u_n), \qquad j=\overline{1,n},
\end{equation}
where $\tau_j, d_j, f_j $ -- certain parameters and nonlinearities of the RD
system correspondingly. In this case Grunwald-Letnikov scheme can be
represented as
\begin{equation*}
u_{j,i}^{k}-\frac{d_j (\Delta t)^{\alpha_j}}{\tau_j (\Delta x)^2}
\left(u_{j,i-1}^{k}-2u_{j,i}^{k}+u_{j,i+1}^{k} \right)-\frac{(\Delta
t)^{\alpha_j}}{\tau_j}f_j(u_{1,i}^{k},...,u_{n,i}^{k})=
\end{equation*}
\begin{equation*}
=(\Delta t)^{\alpha_j}\sum \limits_{p=0}^{m}\frac{(k \Delta t)^{p-\alpha_j}}{%
\Gamma(p-\alpha_j+1)}\frac{\partial^p}{\partial t^p} u_{j,i}^{0}-\sum
\limits_{l=1}^{k} c_{l}^{(\alpha_j)}u_{j,i}^{k-l},
\end{equation*}
\begin{equation*}
c_{0}^{(\alpha_j)}=1, \qquad c_{l}^{(\alpha_j)}=c_{l-1}^{(\alpha_j)}\left(1-%
\frac{1+\alpha_j}{l}\right), \qquad l=1,2,...
\end{equation*}
where $u_{j,i}^{k} \equiv u_j(x_i,t_k) \equiv u_j(i \Delta x, k \Delta t),
\quad m=[\alpha]$.

The applied numerical schemes are implicit, and for each time
layer they are presented as the system of algebraic equations
solved by Newton-Raphson technique. Such approach makes it
possible to get the system of equations with band Jacobian for
each node and to use the sweep method for the solution of linear
algebraic equations. Calculating the values of the spatial
derivatives and corresponding nonlinear terms on the previous
layer, we obtained explicit schemes for integration. Despite the
fact that these algorithms are quite simple, they are very
sensitive and require small steps of integration, and they often
do not allow to find numerical results. In contrast, the implicit
schemes, in certain sense, are similar to the implicit Euler's
method, and they have shown very good behavior at the modeling of
fractional reaction-diffusion systems for different step size of
integration, as well as for nonlinear function and the power
function of fractional index.

We have considered here the kinetics of formation of dissipative structures
for different values of $\alpha $. These results are presented on Fig. %
\ref{rys4} and \ref{rys5}.
\begin{figure}[tbp]
\begin{center}
\includegraphics[width=1.0\textwidth]{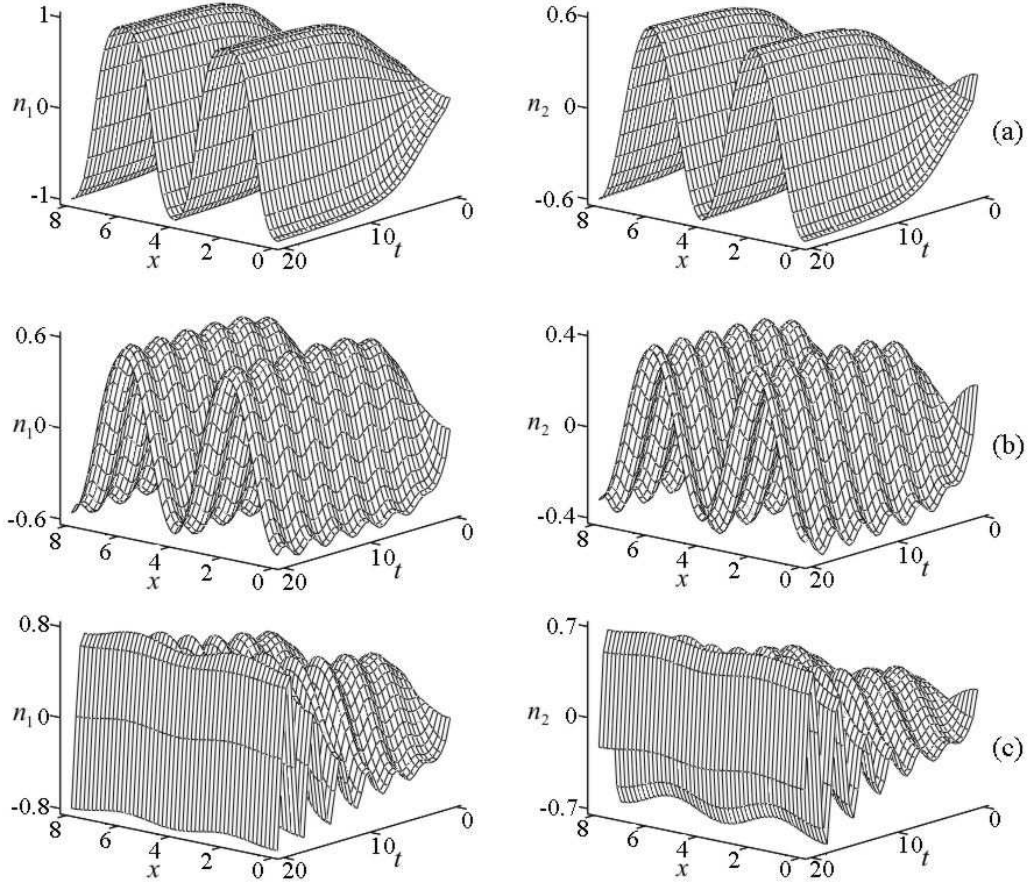}
\end{center}
\caption{Numerical solution of the fractional reaction-diffusion equations (%
\protect\ref{1}),(\protect\ref{2}). Dynamics of variable $n_1$
(left column)
and $n_2$ (right column) on the time interval (0,20) for $\alpha=0.8, l_x=8, \mathcal{A}%
=-0.1, \protect\beta=2, \tau_2=1, l^{2}=0.05, L^{2}=1$; $\quad$ $\protect\tau _{1}/
\protect\tau _{2}=0.75$ -- (a), $\protect\tau _{1}/ \protect\tau
_{2}=0.37$ --  (b), $\protect\tau _{1}/ \protect\tau _{2}=0.3$ --
(c). Initial conditions are: $n_1^0=\overline{n}_{1}-0.05cos(k_0x),
n_2^0=\overline{n}_{2}$.} \label{rys4}
\end{figure}

\begin{figure}[tbp]
\begin{center}
\includegraphics[width=1.0\textwidth]{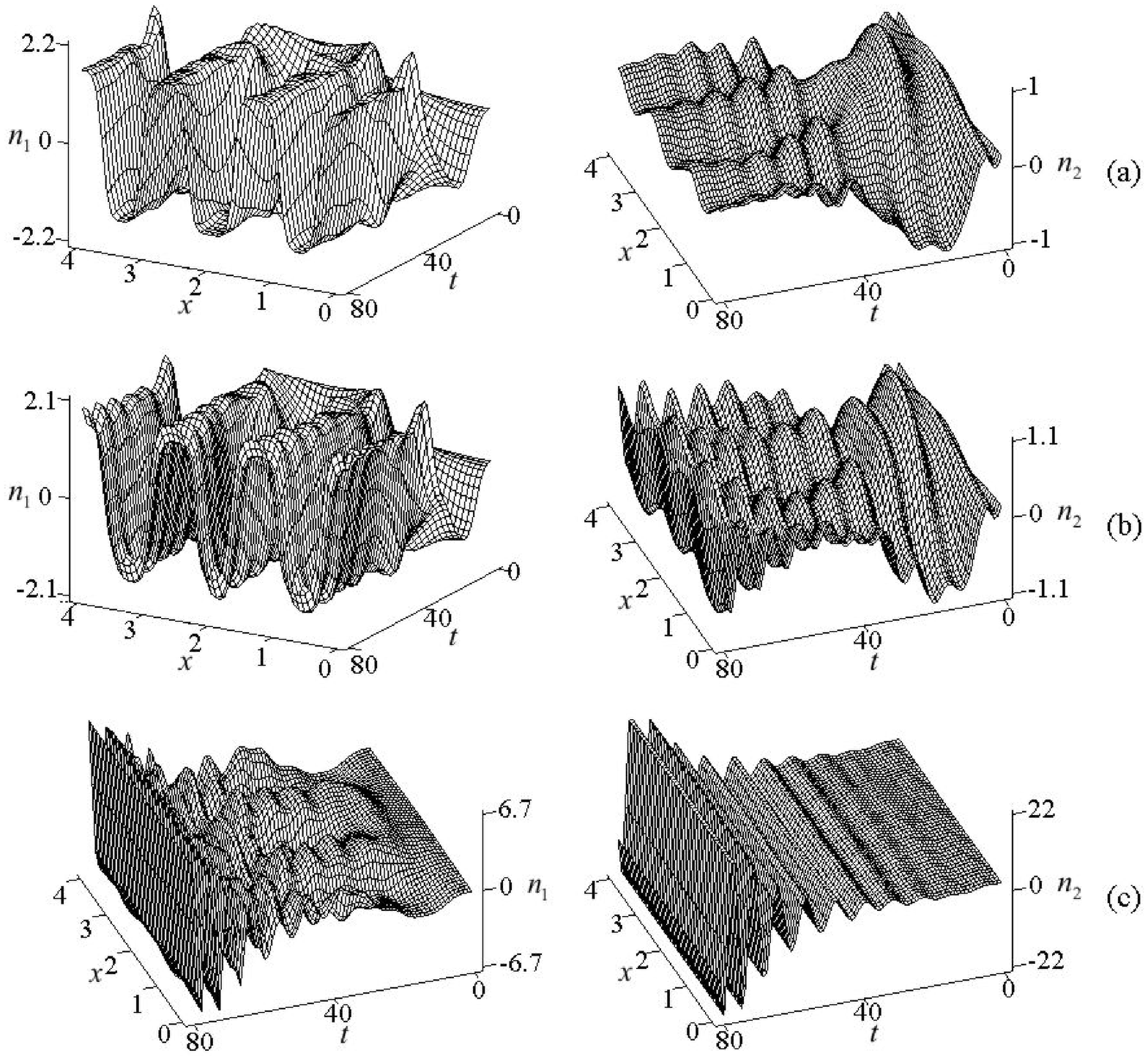}
\end{center}
\caption{Numerical solution of the fractional reaction-diffusion equations (%
\protect\ref{1}),(\protect\ref{2}). Dynamics of variable $n_1$
(left column)
and $n_2$ (right column) on the time interval (0,80) for $\alpha=1.8, l_x=4, \mathcal{A}%
=-0.1, \protect\beta=2, \tau_2=1, l^{2}=0.05, L^{2}=1$; $\quad \protect\tau
_{1}/ \protect\tau _{2}=10$ -- (a), $\protect\tau _{1}/
\protect\tau _{2}=7$ -- (b), $\protect\tau _{1}/ \protect\tau
_{2}=6$ -- (c). Initial conditions are:
$n_1^0=\overline{n}_{1}-0.05cos(k_0x), n_2^0=\overline{n}_{2}$. }
\label{rys5}
\end{figure}

The simulations were carried out for a one-dimensional system on an
equidistant grid with spatial step $h$ changing from 0.01 to 0.1 and time
step $\Delta t$ changing from 0.001 to 0.1. We used imposed Neuman (\ref{bc1}%
)\ or periodic boundary conditions (\ref{bc2}). As the initial condition, we
used the uniform state which was superposed with a small spatially
inhomogeneous perturbation.

It is very well known that in the case of  $\alpha =1$ at $l/L\ll
1,trF<0,\det F<0$\ the homogeneous distribution (\ref{e}) is unstable
according to wave number \cite{pr,ch,m90,KO,dk89,dk03,lg,gl,sbg,mo02}
\begin{equation*}
k=k_{0}=(a_{11}a_{22}-a_{12}a_{21})^{1/4}(Ll)^{-1/2}
\end{equation*}%
As a result at certain parameter $\mathcal{A}$ such that $\overline{n}_{1}\in (-1,1)$
the system can be  unstable and the steady state solutions in the form of
nonhomogeneous dissipative structures arise. Such type systems have rich
dynamics, including steady state dissipative structures, homogeneous and
nonhomogeneous oscillations, and spatiotemporal patterns. In this paper, we
focus mainly on the study of general properties of the solutions depending
on the value of $\alpha $ and the ratio of the characteristic times $\tau_1/\tau_2$.

Fig. \ref{rys4} (a)-(c) show evolution of the dissipative
structure formation for $\alpha < 1$. At certain values of $\tau_1/\tau_2$
we have steady state solution -- (a), then with decreasing ratio
of $\protect\tau _{1}/ \protect\tau _{2}$ nonhomogeneous
pulsating structures -- (b) and transformation of this oscillating
regime to  homogeneous oscillations -- (c).

The evolutionary dynamics for $\alpha>1$ as the  value of $\protect\tau _{1}/
\protect\tau _{2}$ decreases is shown on the Fig.
\ref{rys5}(a)-(c). First plot corresponds to the steady state
structures -- (a) which start to oscillate at smaller  ratio of
characteristic times and eventually homogeneous oscillation are
taking place in FRD system -- (c). Such behavior is due to the case that
the oscillatory perturbations are damping in the first situation,
and then small oscillations are steady state.
With the further decreasing of the value $\protect\tau _{1}/ \protect\tau _{2}$
the steady state oscillations are unstable and dynamics changes to the homogenous temporary
behavior (Fig. \ref{rys5}(c)).

The emergence of homogeneous oscillations, which destroy pattern formation
(Fig. \ref{rys5}(a)-(c))\ \, has deep physical meaning. The matter is that
the stationary dissipative structures consist of smooth and sharp regions of
variable $n_{1}$, and the smooth shape of $n_{2}$. The linear system
analysis shows that the homogeneous distribution of the variables is
unstable according to oscillatory perturbations inside the wide interval of $%
\overline{n}_{1},$ which is much wider then interval ($-1,1$). At the same
time, smooth distributions at the maximum and minimum values of $n_{1}$ are $%
\pm \sqrt{3}$ correspondingly. In the first approximation, these smooth
regions of the dissipative structures resemble homogeneous ones and are
located inside the instability regions. As a result, the unstable
fluctuations lead to homogeneous oscillations, and the dissipative
structures destroy themselves. We can conclude that oscillatory modes in
such type FODEs have a much wider attraction region than the corresponding
region of the dissipative structures.

It should be noted that the pulsation phenomena of the dissipative
structures is closely related to the oscillation solutions of the ODE
(Fig. 3). Moreover, the fractional derivative of the first variable has
the most impact on the oscillations emergence. It can be obtained by
performing a simulation where the first variable is a fractional derivative
and the second one is an integer. It should be emphasized that the
distribution of $n_{2}$, within the solution, only shows a small deviation
from the stationary value (that is why this variable is not represented in
the figures).

\section{Conclusion}

In this article we consider possible solutions of reaction
diffusion system with fractional derivatives. Special attention is
paid to FODEs linear theory of instability which is analyzed
in detail. It was shown that three parameters: fractional
derivative index $\alpha $ the ratio of the
characteristic times $\tau _{1}/\tau _{2},$ and homogeneous solution $%
\overline{n}_{1}$ determine 3-d marginal surface. Inside this
surface the system is unstable and outside it is stable. Nonlinear
dynamics of FODEs is investigated by computer simulation of the
characteristic examples.

By the computer simulation of the fractional reaction-diffusion
systems we provided evidence that pattern formation in the
fractional case, at $\alpha $ less than a certain value, is
practically the same as in the regular case scenario $\alpha =1$.
At $\alpha >{\alpha }_{0}$, the kinetics of formation becomes
oscillatory. At $\alpha ={\alpha }_{0}$, the oscillatory mode
arises and can lead to nonhomogeneous or homogeneous oscillations.

\end{document}